# Ultra-wideband optical coherence elastography from acoustic to ultrasonic frequencies

Xu Feng,[a,b,1] Guo-Yang Li,[a,b,1] and Seok-Hyun Yun[a,b,c,]*

[a] Wellman Center for Photomedicine, Massachusetts General Hospital, 50 Blossom St. BAR-8, Boston, Massachusetts 02114, USA

[b] Department of Dermatology, Harvard Medical School, Boston, Massachusetts 02114, USA

[c] Harvard-MIT Health Sciences and Technology, Cambridge, Massachusetts 02139, USA

[1] Co-first authors with equal contribution

* Correspondence to syun@hms.harvard.edu

**ABSTRACT**

Visualizing elastic waves by noninvasive imaging has been useful for analyzing the mechanical properties of materials and tissues. However, the maximum wave frequency of elastography has been limited to ~10 kHz due to the finite sensitivity to small vibration and finite imaging speed. Here, we present an optical coherence elastography technique that extends the frequency range to MHz by noise reduction, anti-aliasing demodulation, and advanced wave analysis. Our system can measure the stiffness of hard (GPa) materials including bones with mm-scale resolution and characterize soft, viscoelastic materials from 100 Hz to 1 MHz. The dispersion of Rayleigh surface waves over the wide frequency range allowed us to profile depth-dependent shear modulus (10 kPa to 100 MPa) in cartilages *ex vivo* and the human skin *in vivo*. This technique opened a new window for the characterization of materials *in situ* with 3-dimensional resolution.

# INTRODUCTION

Measurement of the mechanical properties of materials is routinely performed in many places in sciences, engineering and industries, as well as hospitals[1-4]. Widely used tools include strain-stress testing[5] and dynamic mechanical analysis[6] for measuring bulk properties, and atomic force microscopy (AFM)[7,8] and microrheology[9,10] for local measurements. In clinical medicine, elastography has been adopted for disease diagnosis[11-14]. Elastography allows to noninvasively measure the elasticity of tissues in normal and abnormal states using medical imaging modalities, such as ultrasound and magnetic resonance imaging (MRI). In all these tools, samples under test are deformed by some force, their responses are measured, and the mechanical properties are calculated from the data[15,16]. The timescale or frequency range of the measurement spans from quasi static (as slow as $10^{-5}$ Hz) to acoustic (as fast as $10^3$ Hz) ranges. Higher speeds up to a few tens of kHz have been used in AFM[17] and in our recent work on dynamic optical coherence elastography (OCE)[18]. In the latter, elastic waves of short impulse or continuous-wave monotones are generated, their propagation within a tissue is visualized, and from the data the wave velocities are determined and related to the elastic moduli of the tissue[15,19].

While the material analysis thus far has been focused on the quasi-static to acoustic ranges, we hypothesized that a higher frequency range beyond 10 kHz can offer a window of opportunity that was previously underappreciated especially for elastography. First, the higher frequency data can reveal the viscoelastic characteristics of materials in the shorter time scale. Dynamic mechanical analysis (DMA) is widely used to characterize viscoelasticity, but it has a limited frequency range (1 to 100 Hz). The time-temperature superposition technique can mimic the high-frequency responses of simple, homogeneous materials[20] but is not applicable to composite materials and living matters[9]. Ultrasound non-destructive testing (NDT)[21,22] is an established technique using ultrasound waves, typically at 500 kHz to 20 MHz, to measure the time of flight across a material or structure with well-defined boundaries to determine bulk elastic properties, but this technique is not directly applicable to elastography and especially for complex materials such as tissues. Second, since the spatial resolution of elastography is approximately given by the wavelength of elastic waves[23], the higher frequency can lead to the higher resolution. Third, wave velocity dispersion over an extended frequency range is useful to extract more detailed mechanical information such as the depth-dependent variation of elasticity and internal stress. The depth information is not accessible using laser Doppler vibrometry or AFM-based rheology that only probes the surface[24-26].

Despite all these anticipated benefits mentioned above, it is technically challenging to extend the upper frequency limit. At the same amplitude of vibration, the energy dissipation or damping of a wave increases with $\sim f^3$ ($f$, frequency) because the wave energy is proportional to $f^2$ and the viscosity tends to grow with $f$. So, with increasing frequency, the wave amplitude must be reduced by $\sim f^{3/2}$ to avoid excessive sample heating or damage. Detecting the reduced amplitude requires improved sensitivity, for example, by a factor of a thousand when comparing 100 kHz to 1 kHz. To analyze waves with submicron amplitudes, a nanometer-scale sensitivity is required to an instrument. OCE is an emerging elastography technique[11,27] built on optical coherence tomography (OCT)[28]. OCE has been developed with a variety of different system architectures and applied to various tissue types, such as cornea[18,29], sclera[30], breast[31], brain[32], and skin[33]. As OCE uses optical interferometry, it offers superior detection sensitivity in the order of nanometers, compared to micrometers for ultrasound and millimeters for MRI. The high sensitivity of OCE make it a good candidate for elastography beyond the acoustic range.

Here, we demonstrate an OCE system capable of visualizing elastic waves from the acoustic to ultrasonic frequencies up to a few MHz. We developed a novel aliasing technique, demodulation algorithm, and jitter-correction method to handle such high frequencies far beyond the typical axial line-scan (A-line) rates of OCT. After verifying the system with homogeneous materials, we apply it to dynamic shear analysis of soft materials, and stiffness mapping of complex tissues in a knee joint *ex vivo* and human skin *in vivo* with high, 3-dimensional resolution. Our results demonstrate the benefits and broad applicability of ultra-wideband OCE.

**RESULTS**

**Signal demodulation of high-frequency vibration in swept-source OCE**

Figure 1a depicts a basic interferometer used in swept source OCT, where the optical wavenumber is tuned repeatedly with a period of $T$: $k = k_0 + k_1[t]$, where $[t] = mod(t,T) - T/2$, $k_0 = 2\pi/\lambda$ ($\lambda$, center wavelength), and $k_1$ is a tuning rate[34]. Imagine a mirror-like sample that vibrates with an amplitude $\delta$, angular frequency $\omega_m (= 2\pi f_m)$, and reflection coefficient $r$ at a mean depth of $z_0$. As the depth is modulated by $\delta \cos(\omega_m t)$, we obtain an interferometric signal $I(t) = r \cos(2k_0 z_0 + 2k_1 z_0 t + 2k_0 \delta \cos(\omega_m t) + 2k_1 t \delta \cos(\omega_m t))$, where $2k_1 z_0 T$ and $\omega_m T$ are assumed to be multiples of $2\pi$ for convenience. Figure 1b illustrates $I(t)$ for $\delta \ll \lambda$. For the typical case of $k_1 T \ll k_0$ and introducing $\omega_0 = 2k_1 z_0$, we get $I(t) \approx r \cos(2k_0 z_0 + \omega_0 t + 2k_0 \delta \cos(\omega_m t))$. In the frequency domain, it consists of an amplitude $r J_0(k_0 \delta)$ at the carrier frequency $\omega_0$ and two first harmonic sidelobes at $\omega_0 \pm \omega_m$ with an amplitude of $r J_1(k_0 \delta) \approx r k_0 \delta$. This is depicted in Fig. 1c. In OCT, the frequency-domain analysis is performed by discrete Fourier transform $F(\omega) = FFT\{I(t)\}$ for each A-line data. The frequency resolution is about equal to the half of the A-line rate $f_A = 1/T$. So, the first harmonic sidelobes overlap with the main peak when $f_m < 0.5 f_A$. This condition has always been met in conventional OCE. The carrier and modulated components interfere with each other in the pixel at $\omega_0$ with a phase $\phi \approx 2k_0 z_0 + 2k_0 \delta \cos(\omega_m t)$. From the amplitude of the phase variation, $\delta$ is readily measured. This method has been commonly used in OCE to date.

We realized that the condition $f_m < f_A/2$ is not necessarily required. When $f_m > f_A/2$, the modulation sidelobes separate from the carrier (Fig. 1d). In this case, one cannot directly apply the conventional method. We rewrite $I(t) \xrightarrow{k_0 \delta \ll 1} Re\{re^{i2k_0 z_0}[e^{i\omega_0 t} + k_0 \delta e^{i(\omega_0 - \omega_m)t} + k_0 \delta e^{i(\omega_0 + \omega_m)t}]\}$. Now that the modulation component is separated from the carrier, the vibration amplitude $k_0 \delta$ manifest itself in the amplitude of the Fourier component at $\omega_0 \pm \omega_m$. In this case, $\delta$ can be measured from the amplitude, not the phase. However, this is only possible for the mirror-like sample when there is no other signal (background) present at $\omega_0 \pm \omega_m$. Consider a tissue-like sample that has continuous scattering points along its depth with $r(z)$. The problem of retrieving $\delta$ from the scatterer at $z_0$ becomes more complicated. Now the Fourier component $F(\omega_0 - \omega_m)$ contains not only the vibration signal (negative sideband) originated from the scatterer at $\omega_0$ but also the time-independent reflection signal (carrier signal) from a different scatterer at $z_0 - z_m$, where $z_m = \omega_m/(2k_1)$. Depending on whether $r(z_0 - z_m) > r(z_0)k\delta$ or not, the phasor circle may or may not exclude the origin, and $\delta$ is embedded differently in the phase and amplitude of $F(\omega_0 - \omega_m)$. Furthermore, $F(\omega_0 - \omega_m)$ also contains a vibration signal (positive sideband) originated from another scatterer located at $z_0 - 2z_m$. Figure 1e (i) shows a phasor diagram illustrating this situation.

Our solution to this problem is as follows. In one embodiment, we may add a large constant bias to $F(\omega_0 - \omega_m)$ to move the rotating phase circles far away from the origin and then retrieve $\delta(z_0)$ from the time-varying amplitude of phase after proper renormalization with the bias and $r(z_0) = \langle F(\omega_0) \rangle$, where $\langle \; \rangle$ denotes time average. In another embodiment, the background $\langle F(\omega_0 - \omega_m) \rangle$ is subtracted from $F(\omega_0 - \omega_m)$. This brings the phasor circles to the origin as depicted in Fig. 1e (ii). $\delta(z_0)$ and $\delta(z_0 - 2z_m)$ can then be identified from a Fourier transform of $F - \langle F \rangle$. We used this second algorithm in experiments. Its rigorous mathematical description is provided in Methods. The situation of $f_m > f_A/2$ is equivalent to signal *aliasing* that occurs when the signal frequency is greater than the sampling rate. In fact, our algorithm works for any $f_m$ values regardless of whether it is aliased or not. Thus, it is a general algorithm for OCE.

**Experimental demonstration of anti-aliasing demodulation**

We used a swept-source OCT system previously built using a polygon-scanner wavelength-swept laser providing a center wavelength of 1307 nm, 3-dB bandwidth of 80 nm, axial resolution of 16 $\mu m$, A-line rate $f_A$ of 43.2 kHz, and maximum average optical power of 12 mW on samples (Supplementary Fig. S1). To test our demodulation scheme, we used a piezoelectric transducer (PZT) actuator block as a mirror-like sample and apply a sinusoidal waveform synchronously with a master clock in the OCT system (Supplementary Fig. S2). To generate mirror-like reflection, a flat glass plate was attached to the PZT. The laser power on the sample was attenuated to generate a signal-to-noise (SNR) of 40 dB at the air-glass interface. Fig. 2a shows a typical A-line profile in the Fourier domain when $f_m$ was 679.6 kHz. The signal appeared at the 107[th] pixel. The interval between pixels is 21.5 µm in depth or 52.7 kHz. Two first harmonic sidelobes emerged at about 12 pixels away from the main peak. The side lobes were lower by 20 dB in signal power (10 dB in amplitude) than the main lobe. Since the power ratio is $20 \, log_{10}(k_0\delta)$, we get $\delta$ =20.4 nm. The sideband position increased linearly with the vibration frequency, with the expected slope of 0.019 pixel / kHz (Fig. 2b).

**Noises and time jitter correction for ultrasonic frequencies**

Various noises ultimately limits the system's sensitivity to vibration and the accuracy of wave velocity measurement. The fundamental limit comes from optical SNR, $X$, in A-lines. The length of the modulation-induced phasor is $r(z_0)k_0\delta(z_0)$. The minimum detectable amplitude $\delta_{min}$ is then given by $r(z_0)k_0\delta_{min}(z_0) = r_{min}(z_0 - z_m)$, where $r_{min}(z_0 - z_m) = r_{min}(z_0) = r(z_0)X(z_0)^{-0.5}$ is the minimum detectable reflection coefficient by definition[35]. So, $k_0\delta_{min} = X^{-0.5}$, same for non-aliased and aliased regimes. We note that although $\delta(z_0)$ is obtained from sidebands at $z_0 \pm z_m$, its error is governed by the SNR of the main peak at $z_0$. In OCE, $\delta$ is typically extracted from a set of M-mode data consisting of $N$ A-lines. Then, the sensitivity is improved by $\sqrt{N}$, and $k_0\delta_{min} = X^{-0.5}N^{-0.5}$. For the above experiment, where $N = 108$ and $X = 10^4$, we find $\delta_{min}$ =195 pm. Another fundamental noise source is 1/f electrical noise. SNR at low frequencies may be limited by the 1/f noise rather than the reflectivity-limited phase noise. Mechanical vibrations of optical components add noises. In our system, the mechanical jitters of beam scanners and rotating polygon filters were suppressed to practically acceptable levels.

Additionally, time jitters in electronics boards and signal generators generate phase noise. This noise increases in proportion to the modulation frequency and can become dominant at the ultrasonic range. In our system, we found that the sweep period $T$ has a Gaussian jitter with a standard deviation $\sigma_T$ of 4.9 ns (Supplementary Fig. S3). The time error accumulates through successive sweeps (Fig. 2c), resulting in a phase error $\sigma_L$. Let $\tau_i$ denote the period jitter in the $i$-th sweep, where $\langle \tau_i \rangle = 0$ and $\langle \tau_i - \langle \tau_i \rangle \rangle = \sigma_T$. The timing of the $m$-th A-line is given by $t_m = mT + \sum_{i=1}^{m}\tau_i$. The output from the $m$-th A-line is $F_m = k_0\delta e^{i\omega_m t_m}$. $\delta$ is determined from the inverse discrete Fourier transform of $F_m$ at $\omega_m$. The result is $k_0\delta(1 + i\omega_m \sum_{m=1}^{N}\sum_{i=1}^{m} t_i /N)$. The standard variation of the second term is $\sigma_L \approx \omega_m\sigma_T\sqrt{(N+1)(2N+1)/6N^2}$ (Supplementary Note 1). For $N \gg 1$, we get $\sigma_L \approx \sqrt{N/3}\,\omega_m\sigma_T$. The total phase noise $\Delta\phi$ is a square sum of the SNR- and jitter-induced noises: $\Delta\phi = \sqrt{1/NX^2 + 0.58N(\omega_m\sigma_T)^2}$. To measure the time jitter in real time and correct for it, we recorded the modulation waveform by using another digitizer channel in the system. The waveform was generated by a trigger synchronized with a particular laser wavelength, so it is supposed to contain the same time jitters as in the sweep period. We obtained the phase of the modulation waveform (applied to the PZT) and subtracted it from the phase of the demodulated signal of each M-scan (Supplementary Note 1). This phase correction was effective (Fig. 2d). The total phase noise after the correction coincides with the SNR limited noise (Fig. 2e).

Finally, the phase noise causes uncertainty in determining the wave velocity, $v$, from the phase variation along the propagation distance. The error $\Delta v$ in velocity is given by $\Delta v/v = \Delta\phi/2\pi\,(\lambda/L)$, where $L$ is the effective measurement length. In soft materials, elastic waves are attenuated, and $L$ is typically a couple of wavelengths. Therefore, $\Delta\phi = 0.01$ yields a high accuracy of $\Delta v/v < 1\%$.

**Ultrasonic OCE of hard materials**

To verify our system at ultra-high frequencies, we measured various hard materials. Mechanical waves were excited on the surface by using a custom-made contact probe with the piezoelectric actuator (see Methods). For a 5 mm-thick acrylate plastic block, we obtained a flat dispersion curve over 1 to 2 MHz (Fig. 3a). Since the wavelength (< 1.4 mm) is smaller than the thickness, the excited wave is the Rayleigh surface wave with a velocity $c_R = (0.862 + 1.14\nu)/(1 + \nu)\, c_s$, where $c_s$ is the pure (bulk) shear wave speed and $\nu$ denotes the Poisson's ratio. The shear elastic modulus $\mu$ is given by $\mu = \rho c_s^2$, where $\rho$ is the density. From the measured $c_R$= 1285 ± 16 m/s and known $\rho$= 1.2 g/cm³ and $\nu$= 0.37, we obtained $\mu$= 2.26 ± 0.06 GPa. This value at MHz is slightly higher than the literature value of 1.7 GPa measured by quasi-static mechanical tools[36]. The difference is presumably due to the weak viscoelasticity of the material. Next, we measured a 1 mm-thick polystyrene plate. Figure 3b shows a representative $k$-domain plot measured at 787.6 kHz. The dispersion relations for the A0 and S0 waves were fitted using the Lamb wave model[37]. The result was $\mu$= 1.64 ± 0.05 GPa, in comparison to quasi-static shear modulus of ~ 1.3 GPa reported in literature[38]. We also performed OCE on glass and metal (Supplementary Fig. S4). We obtained $\mu$= 24.0 ± 0.61 GPa for borosilicate glass and 45.7 ± 1.79 GPa for copper (Fig. 3c). These values were consistent with their quasi-static shear moduli[39,40], indicating their near-pure elasticity. We also performed OCE on the cortical surface of a bovine tibia bone specimen ($\rho$= 1.6 g/cm³ and $\nu$= 0.3) and found $\mu$ to vary from 2.36 ± 0.04 GPa at 485 kHz to 3.60 ± 0.12 GPa at 2 MHz (Fig. 3d). This frequency dependence is in part due to the viscoelasticity of the bone but also contributed by its depth-dependent variation of modulus. Later we will show how the wave velocity dispersion can be used to measure depth-dependent elasticity.

**Dynamic shear analysis of uniform soft materials**

Next, we measured soft polymer samples using surface waves excitation in a frequency range of 0.1 kHz to 100 kHz. Figure 4 displays the experimental and analysis data from a rubber (Ecoflex 00-50), polydimethylsiloxane (PDMS), and a tough hydrogel[41], respectively. Figures 4 (a, d, g) show measured surface displacement profiles at 40 kHz. The profiles contain not only the Rayleigh waves but also supershear surface waves[42,43], which is evidenced by the two separated peaks in the wavenumber domain (Supplementary Fig. S5). We isolated the Rayleigh wave and derived the complex wave number $k_r + ik_i$ of the bulk shear wave (Supplementary Fig. S6). The measured speed ($v = \omega_m/k_r$) and the coefficient of attenuation ($k_i$) are shown in Fig. 4 (b, e, h). The complex shear modulus, $\mu' + i\mu''$, is related to the complex wave numbers via:

$$\mu' = \rho\omega_m^2\, (k_r^2 - k_i^2)/[(k_r^2 - k_i^2)^2 - 4k_r^2 k_i^2] \quad (1)$$

$$\mu'' = 2\rho\omega_m^2\, k_r k_i/[(k_r^2 - k_i^2)^2 - 4k_r^2 k_i^2]. \quad (2)$$

The measured storage and loss shear moduli are plotted in Fig 4 (c, f, i). They show characteristic frequency dependencies due to viscoelasticity. For the rubber, $\mu'$ and $\mu''$ increase from 50 to 160 kPa and from 9 to 323 kPa, respectively, over 0.2 kHz to 50 kHz. $\mu''$ grows at a faster rate above 10 kHz and becomes greater than $\mu'$ above 40 kHz. The crossover of $\mu'$ and $\mu''$ causes strong wave attenuation ($k_i/k_r$ = 0.41). When $k_i/k_r > 1$, waves become overdamped, and their velocities cannot be determined. At low frequencies, our data agreed with previous measurements by MRI elastography, which showed $\mu'$ and $\mu''$ to increase from 30 to 80 kPa and from 10 to 40 kPa over 0.2 to 7.8 kHz[44]. However, the dramatic increase of $\mu''$ beyond 10 kHz is only revealed by the ultrasonic OCE. For PDMS, our data up to 130 kHz show no crossover between $\mu'$ and $\mu''$. The viscoelastic modulus of PDMS in the high frequency range was previously estimated only by the time-temperature superposition method using DMA data below 100 Hz[45]. Among the three materials, the tough hydrogel sample showed the lowest viscosity. The elasticity of this type of hydrogel has recently been measured using DMA below 100 Hz[41]. The low viscosity is due to the water that lubricates highly-entangled polymer chains and reduces energy dissipation[46]. Our data reveals the unusually low $\mu''$ over the entire acoustic range, suggesting that this material is good at transmitting sounds and even ultrasounds.

**Depth-resolved stiffness mapping of cartilage**

We applied OCE to bovine cartilage in the knee joint *ex vivo* (Fig. 5a). The firm, connective tissue consists of three layers with different mechanical properties: the noncalcified cartilage (NCC), calcified cartilage (CC) and the bone (Fig. 5b). In our three-layer model, the thicknesses and shear moduli of NCC, CC and bone layer are denoted by $h_i$ and $\mu_i$ ($i$ = 1, 2, 3), respectively. The NCC layer is readily distinguishable in the OCT image, from which we determined $h_1$ = 300 μm. Unfortunately, the lower part of the CC layer and the underlying bone are not visible due to the limited optical penetration depth. We estimated the thickness of the CC layer to be $h_2$ = 2.5 mm from cross-sectional cuts of equivalent samples (see Methods) and assumed $h_3 \gg h_2$. We measured the elastic wave motions over 10 to 500 kHz. Figure 5d shows the dispersion of the Rayleigh waves, which shows several features. Besides a dramatic transition between 20 and 40 kHz, there appears to be a weak transition between 80 and 100 kHz. The Rayleigh surface wave has a 1/e amplitude decay depth equal to, approximately, a half wavelength. Therefore, as the frequency increases, the wave is increasingly confined near the surface, and its speed reflects the average stiffness of the region. The high speed below 20 kHz is due to the hard bone ($\mu_3$ = 3.27 GPa, see Fig. 2d). The first downward transition from 20 to 40 kHz is related to the bone-CC interface as the Rayleigh wave moves away from the bone. And the downward transition from 80 to 100 kHz is attributed to the CC-NCC interface and indicates that $\mu_1 < \mu_2$. We developed a three-layer finite element analysis (FEA) model (see Method) and plotted the best-fit result in Fig. 5c, from which we determined $\mu_1$ = 5.6 ± 0.2 MPa and $\mu_2$ = 13.2 ± 0.9 MPa.

Our demodulation method allowed us to generate subsurface wave maps (Supplementary Fig. S7). Figure 5d shows cross-sectional displacement images obtained at 100, 140, and 500 kHz, respectively. It is apparent that the wavelength is longer in NCC than CC. We performed FEA using the geometrical and elastic parameters obtained from the curve fitting to generate displacement maps. The simulation reproduces the measured displacement fields quite well.

**Depth-resolved stiffness mapping of *in vivo* human skin**

We next examined the stiffness of human fingertip skin *in vivo* over 100 Hz to 500 kHz. We have previously measured the moduli of the epidermis, dermis (D), and hypodermis (H) in the forearm skin by OCE at acoustic frequencies[33]. With ultrasonic frequencies, we were interested in resolving the stratum corneum (SC)[47] and the viable epidermis (VE) layers within the epidermis (Fig. 6a). Figure 6b show cross-sectional displacement images at different frequencies. At 20 kHz, the Rayleigh wave is confined predominantly in the SC and VE layers. Above 300 kHz, the majority of the elastic energy resides in the SC layer.

The wave velocity dispersion curve has a strong frequency dependence in the normal skin condition (Fig. 6c, red circles). To relate the dispersion to shear modulus, let us consider a depth-profile of elastic shear modulus, $\mu(z)$. There are no analytic solutions for wave propagation for arbitrary nonuniform modulus. For small nonuniformity, the wavenumber can be estimated from $k^2 = \rho\omega^2 \int_0^\infty \psi^* \frac{1}{\mu(z)} \psi \, dz$, where $\psi(z)$ is the wave function in a uniform medium with a space-averaged modulus: $|\psi(z)|^2 \approx 2k_z e^{-2k_z z}$, where $k_z \approx 0.31\,k$. We may extend the perturbation theory to large variation with a simplified form:

$$k^2 \approx \frac{\rho\omega^2}{L_z} \int_0^{L_z} \frac{1}{\mu(z)} \, dz \quad (3)$$

where $L_z$ denotes the penetration depth of the Rayleigh wave. We may choose $L_z = 1/2k_z \approx 0.25\,\lambda$ or more generally, $L_z = a\lambda = av/f$, where $a$ is a fitting constant. By differentiating both sides of Eq. (3) with respect to $f$, we obtain

$$\mu(z) \approx \rho v^2 \, (v - fv')/(v + fv') \quad (4)$$

where $z = av/f$, $v = 1.048\,v_R$ is bulk shear velocity obtained from the measured Rayleigh wave velocity $v_R(f)$, and $v' = dv/df$. Figure 6d shows $\mu(z)$ obtained from the experimental data. We find $\mu(z)$ to decay rapidly from 56 MPa at the surface to 100 kPa at the asymptotic depth (using $a$= 0.25). This finding is consistent with the fact that water content decreases exponentially from the VE to the surface of SC[48] and that the elastic modulus decreases exponentially with depth in SC.

We measured the same skin immediately after the fingertip was immersed in warm water for 20 min. The thickness of the SC layer increased from 0.31 mm to 0.41 mm due to swelling (Fig. 6e). The wave velocity at high frequencies decreased significantly (Fig. 6c, yellow dots), indicating a softening of the SC layer. The Rayleigh waves became overdamped at frequencies above 50 kHz due to high viscous damping. Using Eq. (4), we obtained $\mu$= 2.6 MPa at the surface of the hydrated SC.

To verify, we performed FEA simulation by modeling the skin as a four-layer structure. We measured the shear moduli of the dermis ($\mu_2$= 9-17 kPa) and hypodermis ($\mu_3$= 2-3 kPa) layers from wave dispersion below 1 kHz[33]. The shear modulus in the SC layer at the pre-hydrated condition was assumed to have an exponentially decreasing function from $\mu_0$= 50 MPa at the surface to $\mu_1$= 2 MPa at the interface between the SC and VE layers (Supplementary Fig. S8). In the hydrated condition, the stiffness of the swollen SC layer was assumed to be uniform and the same as the VE layer ($\mu_0 = \mu_1 = 2$ MPa). In Fig. 6c we plot the dispersion relations obtained by the FEA simulation, which captures the salient features of the dispersion relations.

## DISCUSSION

We have demonstrated an OCE system capable of visualizing elastic waves within various materials and tissues over a wide range of shear modulus from 1 kPa to 10 GPa and across an unprecedented frequency range from static to a few MHz. The fundamental reflectivity-limited performance of the system provided sufficient SNR to determine wave velocities with high accuracy. The measured speed dispersion over the wide frequency range was critical to extract depth-dependent stiffness from multi-layered tissues commonly found in various tissues. The contact area of the PZT probe was optimized to be less than half the wavelength to maximize acoustic impedance matching from the actuator motion to the elastic waves. Throughout the experiments, the peak wave amplitude ranged from several hundreds of nm at acoustic frequencies to as small as 10 nm at ultrasonic frequencies.

The demodulation scheme was highly effective and allowed the upper frequency limit to be no longer limited by the A-line rate. This relaxes the system requirement. The demodulation scheme should be applicable to any swept source OCT systems regardless of their A-line rates. It should be noted that the aliasing scheme does not work for spectra-domain OCT because signal modulation during the A-line acquisition period is averaged out due to the integrating nature of CCD spectrometers (equivalent detector bandwidth is only $f_A/2$). This is analogous to the phase washout motion artifact in spectral-domain OCT[34]. Swept-source OCT is free from the phase washout as long as $f_m$ is lower than the detector bandwidth, which is typically 1000 times $f_A$.

We envision ultra-wideband OCE to be a powerful tool for mechanical characterizations of soft and hard tissues with high spatio-temporal resolution. For example, ultra-wideband OCE may be useful in capturing the full spectrum of the complex rheology of tissues[49-51] and biomaterials[52,53]. With the capability of 3D high-resolution stiffness mapping, ultra-wideband OCE has the potential to enable novel biomechanics-based clinical diagnostics[54-56]. Besides, the ability to detect up to MHz excitation may be useful to study fast neuronal activities through neuro-mechanical coupling and develop novel therapeutics to treat brain diseases[57]. Furthermore, ultra-wideband OCE may be used to extract mechanical stress in thin structures without prior knowledge of material properties[58]. Finally, the emergence of deep learning offers a new approach to apply the information-rich elastography data to multi-parameter inverse problems[59].

## MATERIALS AND METHODS

**Swept-source OCT signal.** The photodetector current $I(t)$ of the OCT interferometry from a point scatter located at $z_0$ can be expressed as

$$I(t) = rP(t)\cos(2k_0z_0 + 2k_1tz_0), \quad (5)$$

where $r$ denotes the reflectance amplitude, $P(t)$ denotes the optical power profile, $k = k_0 + k_1 t$ denotes the wavenumber of the swept source. The Fourier transformation of the photocurrent acquired during time interval $-T/2 \leq t \leq T/2$ with respect to $\hat{k} = 2k_1 t$, denoted by $F(z)$, is given by

$$F(z_0; z) = \int_{-k_1 T}^{k_1 T} I(\hat{k}) e^{i\hat{k}z} d\hat{k}, \quad (6)$$

where $T$ is the period of wavelength sweep (A line period). We assume that the optical power profile has a Gaussian shape with a full width at half maximum (FWHM) $\sigma T$, i.e., $P(t) = \exp\left[-4\ln 2 \frac{t^2}{(\sigma T)^2}\right]$. Then we get

$$F(z_0; z) = \frac{r}{2} \int_{-k_1 T}^{k_1 T} \exp\left[-4\ln 2 \frac{\hat{k}^2}{(2\sigma k_1 T)^2}\right] \{e^{-i(2k_0 z_0 + \hat{k}z_0)} + \text{c.c.}\} e^{i\hat{k}z} d\hat{k}, \quad (7)$$

where 'c.c.' denotes the complex conjugate. When $\sigma < 1$, we can approximate the integral range to infinity and get

$$F(z_0; z) = \frac{r}{2}\left(\sqrt{\frac{\pi}{\ln 2}} \sigma k_1 T\right)\{F_0(z) + \text{c.c.}\}. \quad (8)$$

where we introduce $F_0(z)$ that is defined as

$$F_0(z) = e^{-4\ln 2 \frac{(z-z_0)^2}{(\sigma_z)^2}} e^{-i2k_0 z_0}. \quad (9)$$

where $\sigma_z \equiv \frac{4\ln 2}{\sigma k_1 T}$ corresponds to the FWHM axial resolution.

Without loss of generality, the displacement of the scatter along the laser beam can be expressed as $\Delta z(t) = \delta g(t)$, where $\delta$ is a constant and $g(t)$ is a dimensionless function of time. Then the location of the scatter is $z_0 + \Delta z$. Since the laser is periodically tuned, we introduce $\hat{t} = t - mT$ for the $m$-th A line acquired during $mT - T/2 \leq t \leq mT + T/2$, and $g_m(\hat{t}) = g(\hat{t} + mT)$. Substitution of $z_0$ and $P(t)$ in Eq. (5) with $z_0 + \delta g_m(\hat{t})$ and $P(\hat{t})$, we can get

$$I_m(\hat{t}) = rP(\hat{t})\cos\left(2k_0 z_0 + 2k_1\hat{t}z_0 + 2k_0 A_{z_0} g_m(\hat{t}) + 2k_1\hat{t}\delta g_m(\hat{t})\right). \quad (10)$$

We make assumption that the wavelength tuning range is a small fraction of the mean wavelength, $k_1\hat{t} \ll k_0$, to get

$$\begin{aligned} I_m(\hat{t}) &\approx rP(\hat{t})\cos(2k_0 z_0 + 2k_1\hat{t}z_0 + 2k_0\delta g_m(\hat{t})) \\ &= rP(\hat{t})\{e^{-i(2k_0 z_0 + 2k_1\hat{t}z_0 + 2k_0\delta g_m(\hat{t}))} + \text{c.c.}\} \end{aligned} \quad (11)$$

$$= rP(\hat{t})\left\{e^{-i(2k_0z_0+2k_1\hat{t}z_0)}\left[1+(-2ik_0\delta)g_m(\hat{t})+\tfrac{1}{2}(-2ik_0\delta)^2 g_m^2(\hat{t})+\cdots\right]+\text{c.c.}\right\}.$$

We define

$$G_m^{(1)}(z) = \int_{-\infty}^{+\infty} g_m(\hat{t})e^{i\hat{k}z}\,\mathrm{d}\hat{k},$$

$$G_m^{(2)}(z) = \int_{-\infty}^{+\infty} g_m^2(\hat{t})e^{i\hat{k}z}\,\mathrm{d}\hat{k}.$$

(12)

The Fourier transform of the photocurrent $I(\hat{t})$ is

$$F_m(z_0;z) \approx \tfrac{r}{2}\left(\sqrt{\tfrac{\pi}{\ln 2}}\sigma k_1 T\right)\left\{\left[F_0(z)+(-2ik_0\delta)F_0(z)*G_m^{(1)}(z)+\tfrac{1}{2}(-2ik_0\delta)^2 F_0(z)* G_m^{(1)}(z)+\cdots\right]+\text{c.c.}\right\},$$

(13)

where '*' denotes the convolution.

For harmonic vibration studied here, $\Delta z = \delta \sin(\omega_m t + \varphi)$, where $\delta$, $\varphi$, and $\omega_m$ denote the vibration amplitude, initial phase, and angular frequency, respectively. Then have $g_m(\hat{t}) = \sin(\omega_m \hat{t} + m\omega_m T + \varphi)$, and

$$G_m^{(1)}(z) = \tfrac{1}{2i}\left[e^{i(m\omega_m T+\varphi)}\delta\left(z+\tfrac{\omega_m}{2k_1}\right)-e^{-i(m\omega_m T+\varphi)}\delta\left(z-\tfrac{\omega_m}{2k_1}\right)\right],$$

$$G_m^{(2)}(z) = \tfrac{1}{2}\left[\delta(z)-\tfrac{1}{2}\left(e^{2i(m\omega_m T+\varphi)}\delta\left(z+\tfrac{\omega_m}{k_1}\right)+e^{-2i(m\omega_m T+\varphi)}\delta\left(z-\tfrac{\omega_m}{k_1}\right)\right)\right],$$

(14)

where $\delta(z)$ is the Dirac delta function. Inserting Eq. (13) we get

$$F_m(z_0;z) \approx \tfrac{r}{2}\left(\sqrt{\tfrac{\pi}{\ln 2}}\sigma k_1 T\right)\left\{\left[(1-(k_0\delta)^2)F_0(z)-(k_0\delta)\left(e^{i(m\omega_m T+\varphi)}F_0\left(z+\tfrac{\omega_m}{2k_1}\right)-e^{-i(m\omega_m T+\varphi)}F_0\left(z-\tfrac{\omega_m}{2k_1}\right)\right)+\tfrac{1}{2}(k_0\delta)^2\left(e^{2i(m\omega_m T+\varphi)}F_0\left(z+\tfrac{\omega_m}{k_1}\right)+e^{-2i(m\omega_m T+\varphi)}F_0\left(z-\tfrac{\omega_m}{k_1}\right)\right)+\cdots\right]+\text{c.c.}\right\}.$$

(15)

Taking the first order approximation, we have

$$F_m(z_0;z) \approx \tfrac{r}{2}\left(\sqrt{\tfrac{\pi}{\ln 2}}\sigma k_1 T\right)\left\{\left[F_0(z)-(k_0\delta)\left(e^{i(m\omega_m T+\varphi)}F_0\left(z+\tfrac{\omega_m}{2k_1}\right)-e^{-i(m\omega_m T+\varphi)}F_0\left(z-\tfrac{\omega_m}{2k_1}\right)\right)\right]+\text{c.c.}\right\}.$$

(16)

**Aliasing of the vibration signal.** To derive the vibration signal from the OCT measurement, we consider one of the side lobes in Eq. (16), $F_m\left(z_0; z_0 - \frac{\omega_m}{2k_1}\right)$, and find

$$F_m\left(z_0; z_0 - \frac{\omega_m}{2k_1}\right) \propto (k_0\delta)e^{-i2k_0z_0}e^{im\omega_m T}e^{i\varphi}, \tag{17}$$

which retains the vibration signal. Denote the frequency $f_m = \frac{\omega_m}{2\pi}$ and the A line rate $f_A = \frac{1}{T}$. When $f_m > 0.5f_A$, Eq. (17) gives rise to the aliasing effect. Denote $f_m = \bar{f}_m + 0.5nf_A$, where $n$ is an integer to make $0 \leq \bar{f}_m < 0.5f_A$. Then,

$$F_m\left(z_0; z_0 - \frac{\omega_m}{2k_1}\right) \propto (k_0\delta)e^{-i2k_0z_0}e^{i2\pi m\bar{f}_m T}e^{i\varphi}, \tag{18}$$

when $n$ is an even number, and we can get the apparent frequency $\bar{f}_m$.

$$F_m\left(z_0; z_0 - \frac{\omega_m}{2k_1}\right) \propto (k_0\delta)e^{-i2k_0z_0}e^{i2\pi m(\bar{f}_m - 0.5f_A)T}e^{i\varphi}, \tag{19}$$

when $n$ is an odd number and we get the apparent frequency $(\bar{f}_m - 0.5f_A)$.

**Demodulation of vibrations from internal scatters**

Given the axial coordinate $z_0$, without loss of generality, the Fourier transform of the photocurrent for the $m$-th scan, denoted by $\mathcal{F}_m(z_0)$, is

$$\mathcal{F}_m(z_0) \approx F_m(z_0; z_0) + F_m(z_0 + z_m; z_0) + F_m(z_0 - z_m; z_0), \tag{20}$$

where $z_m = \frac{\omega_m}{2\kappa_1}$. The latter two, comes from the contributions of the sidelobes of two scatters located at $z_0 + z_m$ and $z_0 - z_m$ (denoted by S' and S''), respectively. As the sidelobes contain the vibration signals, we expect to extract the vibrations of S' and S'' from $\mathcal{F}_m(z)$.

Note that the sum $\sum_{m=1}^{N} F_m(z_0 \pm z_m; z_0) = 0$, given that $Nf_m/f_A$ is an integer, where $N$ is the number of A lines. We have $\frac{1}{N}\sum_{m=1}^{N} \mathcal{F}_m(z_0) = F_m(z_0; z_0)$, the static OCT signal of the scatter located at $z_0$. Subtracting this term from $\mathcal{F}_m(z_0)$ we can get,

$$F_m(z_0 + z_m; z_0) + F_m(z_0 - z_m; z_0) \approx \mathcal{F}_m(z_0) - \frac{1}{N}\sum_{m=1}^{N} \mathcal{F}_m(z_0). \tag{21}$$

The left side of Eq. (21) contains the left sidelobe of the scatter S' and the right sidelobe of scatter S''. Here we show the extraction of the vibration for S' as an example. Note that $\frac{1}{N}\sum_{m=1}^{N} \mathcal{F}_m(z_0 + z_m) = F_m(z_0 + z_m; z_0 + z_m)$, the static OCT signal of S'.

$$\left(k_0\delta_{z_0+z_m}\right)e^{i(m\omega_m T+\varphi)} + \frac{F_m(z_0-z_m;z_0)}{\frac{1}{m}\sum_{m=1}^{M}\mathcal{F}_m(z_0+z_m)} \approx \frac{\mathcal{F}_m(z_0) - \frac{1}{N}\sum_{m=1}^{N}\mathcal{F}_m(z_0)}{\frac{1}{N}\sum_{m=1}^{N}\mathcal{F}_m(z_0+z_m)}. \tag{22}$$

where $\delta_{z_0+z_m}$ denotes the vibration amplitude at $z_0 + z_m$. The first term of the left side in Eq. (22) gives the vibration of S'. To eliminate the second term that results from the contribution of the right sidelobe of S'', we note that the phase change of the right sidelobe has an opposite direction as the left side lobe. The two terms can be separated in the frequency domain. Then the last step to extract the vibration of S' is to perform Fourier transform to right side of Eq. (22). An example of the demodulation algorithm is provided in Supplementary Fig. S1.

**Experimental setup.** Our experimental setup (Supplementary Fig. S2) is based on a home-built, swept source OCT system[60]. The system uses a rotating-polygon-mirror wavelength-swept laser[61] with a central wavelength of 1307 nm and a 3-dB bandwidth of 80 nm at an A-line rate of 43.2 kHz. The axial resolution is 16 $\mu$m. The optical beam is scanned by a pair of galvanometer mirror scanners and focused by a wide-aperture scan lens (Thorlabs, LSM54–1310) yielding a long working distance of 64 mm and a transverse resolution of ~30 μm. The average optical power on the sample is 12 mW. A fiber Bragg grating (FBG) and photodiode (PD) provide a pulse signal synched to each wavelength sweep cycle of the laser output. This FBG optical clock ensures time synchronization among the modulation waveform to PZT, OCT beam position scan, and OCT data acquisition. The detector signal was digitized by a data acquisition board (Signatec, PX14400) at a sampling rate of 108 MHz. A total 2048 data points were acquired during each A-line. An input/output (I/O) board (National Instruments, USB-6353) is used to generate analog waveforms for the galvanometer scanners. The I/O board or a function generator (Tektronix, AFG3021C) is used to generate stimulus waveforms. The waveforms are smoothed with a reconstruction filter (Thorlabs, EF122) before feeding to the PZT. During data acquisition, the stimulus waveforms applied to the PZT are simultaneously recorded by the I/O board, which was later used for time jitter correction. Figure S2 depicts the timing diagram for the data acquisition protocol. The M-B scan protocol comprises the acquisition of $M$ consecutive A-lines at each transverse location. The OCT beam is then moved to the next position, and the stimulus waveform is repeated. A total of 96 positions are acquired. Typical, we set $M$ to be in the range of 100 to 250. For $M$ = 108, a single M-scan at each transverse location takes 2.5 ms, and the total measurement time per frequency is 0.24 s.

The harmonic waveforms are sent to a wideband PZT (Thorlabs, PA4CEW) via an amplifier (PiezoDrive, PDm200B for frequencies below 200 kHz or E&I 1020L above 200 kHz). For the ultrasonic OCE of hard materials, the PZT was in direct contact with the sample with a contact area of 2 mm×2 mm to generate sufficient push force. For the dynamic shear analysis of soft uniform materials, a custom mechanical actuator comprising of a PZT and a 3D-printed cylindrical tip with 0.6 mm radius was used. The tip was designed to facilitate the theoretical model of near-field surface wave[62]. For the depth-resolved stiffness mapping of the joint and skin, a 3D-printed prism-shape tip with a line contact of 2 mm is used to reduce wave attenuation and suppress supershear surface wave[33].

**Displacement field analysis.** All algorithms were developed in MATLAB (MathWorks Inc., MATLAB 2019). The raw data collected from the M-B scan was processed using the standard swept-source phase-stabilized algorithm to obtain the complex-valued OCT tomogram[60]. When $f_m > f_A/2$, the correct vibrations occurred at two side lobes. We typically selected the left side lobe, and then applied the demodulated algorithm to extract the surface displacement profiles. Next, we performed a 1-dimensional Fourier transform to move the data from time $t$ domain to frequency $f$ domain. The frequency domain data was filtered at the driven frequency (when $f_m < f_A/2$) or the aliased frequency (when $f_m > f_A/2$) to obtain lower noise waveforms. After we obtained the displacement profiles over the $x$ coordinate, we performed another 1-dimensional Fourier transform to move the data from the spatial $x$ domain to the wavenumber $k$ domain. The wavenumber $k$ of the surface wave was then determined from the plot by selecting the peak corresponding to the Rayleigh surface wave. This filtering in the $k$-$x$ domain is critical to remove other higher-order modes especially at high frequencies. The phase velocity is then given by $v = 2\pi/k$. To obtain the cross-sectional displacement image, the same method used for obtaining the surface wave displacement was applied throughout the entire pixels along the depth $z$. In addition, the motion artefact caused by the surface wave motion were corrected within the sample[63].

**Measure complex wave number from surface waves.** For soft materials, the near-field displacement is primarily dominated by the Rayleigh and supershear surface waves in high frequency regime. An analytical approximation solution for the surface waves excited by a cylindrical probe can be obtained. The surface displacement at lateral coordinate $x$ is[62]

$$u(x) = i\pi \frac{ap_0 K^4}{\rho \omega_m^2} \left[ \frac{J_1(K_R a) K_R}{\mathcal{F}'(K_R)} H_0^{(1)}(-K_R x) + \frac{J_1(K_{SS} a) K_{SS}}{\mathcal{F}'(K_{SS})} H_0^{(1)}(-K_{SS} x) \right], \tag{23}$$

where $J_1$ is the Bessel function of the first kind of order 1, $H_0^{(1)}$ is the Hankel function of the first kind, $a$ is the radius of the cylindrical probe, $p_0$ is the pressure amplitude exerted to the sample by harmonic vibrations of the probe. $\omega_m$ is the vibration frequency. $K$, $K_R$ and $K_{SS}$ are wavenumbers of the shear, Rayleigh surface, and supershear surface waves, respectively. $K_R$ and $K_{SS}$ are roots of the secular equation $\mathcal{F}(\kappa) = (2\kappa^2 - K^2)^2 - 4\kappa^2 \sqrt{\kappa^2(\kappa^2 - K^2)} \, \text{sign}\{\text{Re}(\kappa^2 - K^2)\} = 0$. So, we have $K_R = 1.047K$ and $K_{SS} = (0.4696 - 0.1355i)K$. $\mathcal{F}'(\kappa) = d\mathcal{F}/d\kappa$. To derive $K$ from the experiments, we fit the real and imaginary displacements simultaneously with Eq. (23) using the least-squares method. Representative fitting results can be found in Supplementary Figure S7.

**Finite element analysis.** Finite element analysis (FEA) was performed with Abaqus 6.12 (Dassault Systèmes). We used plane strain models and performed time domain simulations. A localized harmonic surface pressure was applied to excite elastic waves, simulating the PZT actuator in the experiments. Other sides of the model were completely constrained. The geometry and total time of the model was scaled by the wavelength and stimulus frequency $f_m$, respectively. The time increment and total time were $0.1/f_m$ and $20/f_m$, respectively. The width and height of the model were about 16 folds of the maximum wavelength to avoid reflection from the boundaries. The model was divided into uniform layers, each with different thickness and elastic parameters. To model a layer with a gradient property, we assigned a temperature-dependent elastic modulus to the layer and then prescribed the temperature field with an analytical distribution. The gradient mesh was adopted to reduce the computation costs. At the surface the mesh size was smaller than one tenth of the wavelength and at the bottom the mesh size was about half of the wavelength. The element type used in this study was 8-node biquadratic element (CPE8RH).

The convergence of the model was checked by making sure the results independent of the mesh size.

**Preparation of hard materials.** The acrylate plastic block (McMaster-Carr) presented in Fig. 3a had a thickness of 5 mm and an area of 50 mm×50 mm. The polystyrene petri dish (Fisher Scientific) presented in Fig. 3b had a bottom thickness of 1 mm and a diameter of 150 mm. The borosilicate glass coverslip (Fisher Scientific) measured in Supplementary Fig. S5a had a thickness of 0.15 mm and an area of 24 mm×50 mm. The copper foil (All Foils Inc.) measured in Supplementary Fig. S5b had a thickness of 0.3 mm and an area of 15 mm×40 mm. The glass coverslip and the copper foil were placed on a lens holder so that the sample was bounded by the air on two sides.

**Preparation of soft materials.** The bulk silicone rubber has a diameter and height of 60 mm and 12 mm, respectively. It was prepared from Ecoflex 0050 material (Smooth-On Inc) by mixing the Ecoflex 1A and 1B at 1:1 ratio by weight. The mixture was poured into a mold and cured at room temperature overnight. The material was then post-cured in an oven at 80ºC for 2 hours. The bulk PDMS sample has a diameter and height of 120 mm and 50 mm, respectively. It was prepared by using a 2:1 mixing ratio of base elastomer and curing agent (Sylgard 184, Dow Corning) and cured for 45 mins at 85 °C. The hydrogel sheet had a size of 40 mm×40 mm and a thickness of 1.3 mm under fully swollen, highly entangled condition[41]. All soft materials were assumed to have a mass density of $\rho \simeq 1000$ kg/m$^3$.

**Cartilage tissues.** Two fresh bovine tibia bones from two juvenile calves were obtained 1 h post-mortem (Research 87 Inc., Boylston, MA). The tissues were wrapped in a wet towel to keep them well-hydrated until use. Three measurements were performed on the cortical surface of the bone. Four measurements were performed on the femoral condyle region covered by articular cartilage. To measure the thickness of calcified cartilage layer, we made a cross-sectional cut after the OCE experiments.

**Human subject.** We measured the skin of the left index finger on a 31-year-old male subject. The study was conducted at the Massachusetts General Hospital (MGH) following approval from the Institutional Review Board (IRB) of Massachusetts General Hospital and the Mass General Brigham Human Research Office. Written informed consent was obtained from both subjects prior to the measurement and all measurements were performed in accordance with the principles of the Declaration of Helsinki. The measurement site was marked by a surgical grade skin marker. For the hydration test, the whole index finger was immersed in warm water for 20 min, wiped with a clean towel, and then measured immediately.

## Acknowledgements

This study was supported by funding from the National Institutes of Health via grants R01-EY027653, R01-EY033356, and R01-HL098028. The authors thank Dr. Guogao Zhang for providing the hydrogel sample.

## Author contributions

S.H.Y. conceived the idea. G.Y.L. and S.H.Y. developed the theory and signal processing. X.F. and G.Y.L. carried out the experiments and analyzed the results. All authors wrote and reviewed the manuscript.

**Figures**

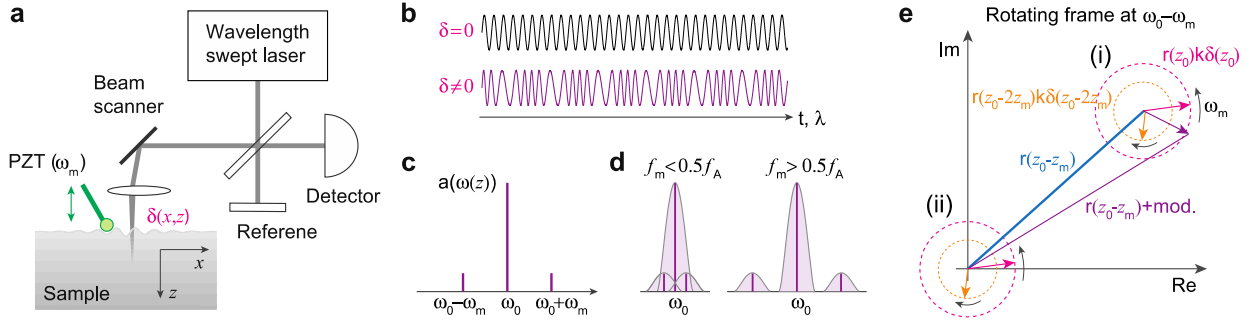

**Fig. 1 Principle of high-frequency OCE. a**, Schematic of a swept-source interferometric setup. A piezoelectric actuator excites elastic waves with angular frequency $\omega_m$ and amplitude profile $\delta(x,z)$. **b**, Example of time-domain detector signals without and with vibration of a mirror. The primary carrier frequency is given by the depth location ($z$) of the reflector and the wavelength sweep speed. **c**, Frequency-domain transform of the trace in b. **d**, Overlap of the carrier and modulation components for low and high frequencies compared to the A-line rate ($f_A$). **e**, Graphical relationship between $\delta(z)$ and $r(z)$ in the Fourier-domain at $\omega_0 - \omega_m$. Depending on whether $r(z_0 - z_m)$ is greater or smaller than the magnitude of rotating phasor $r(z_0)k_0\delta(z_0)$, the pixel phase undergoes an oscillatory or diverging pattern, respectively. For demodulation, the rotating phase circles (i) is brought to the origin (ii) by subtracting the offset $r(z_0 - z_m)$, and $\delta(z_0)$ is determined from the magnitude of the counterclockwise rotating phasor.

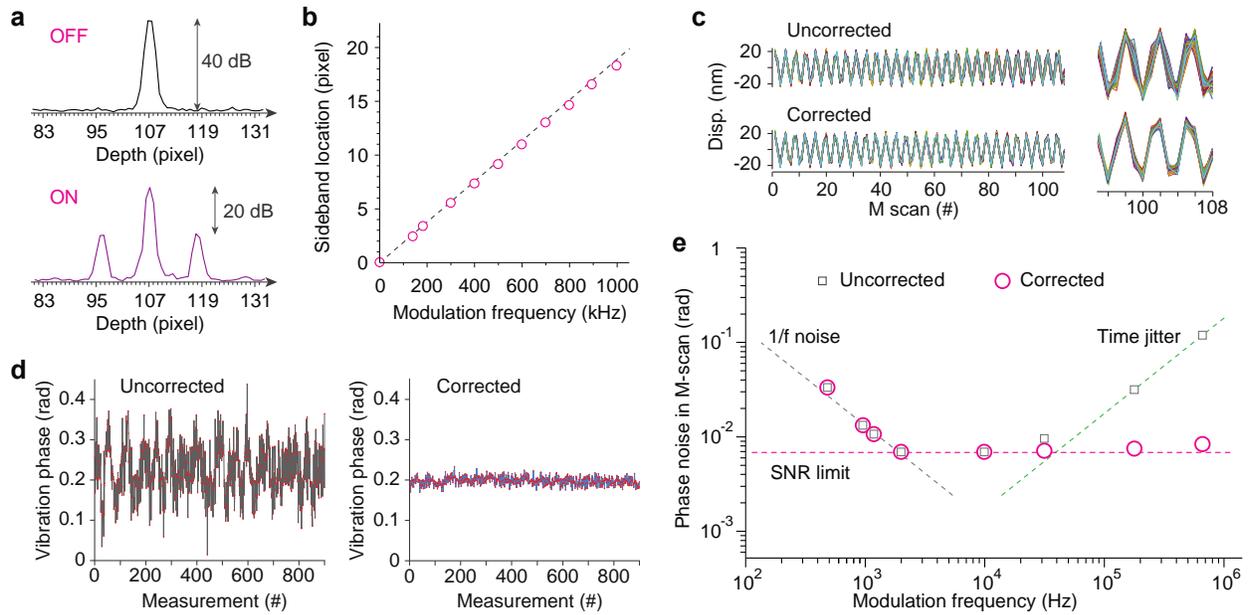

**Fig. 2 Implementation of aliasing in OCE. a**, Signal from the surface of a stationary PZT in the frequency (depth) domain without modulation (top) and with modulation at 679.6 kHz (bottom). **b**, Measured position of modulation peak at different modulation frequency. **c**, Several M-scan traces at 679.6 kHz before and after time jitter correction. **d**, Phase data of the M-scan traces showing noises before and after jitter correction. **e**, Phase noise performance measured with optical SNR of 40 dB and $N$ = 108. The SNR-limited phase noise corresponds to -94 dB/Hz. The 1/f noise is also seen. Note that this graph is for 40 dB SNR. For lower SNR, the optical phase noise level would increase. For example, for 20 dB optical SNR, the optical noise level is moved up to 0.07 rad, crossing the 1/f noise line at 200 Hz.

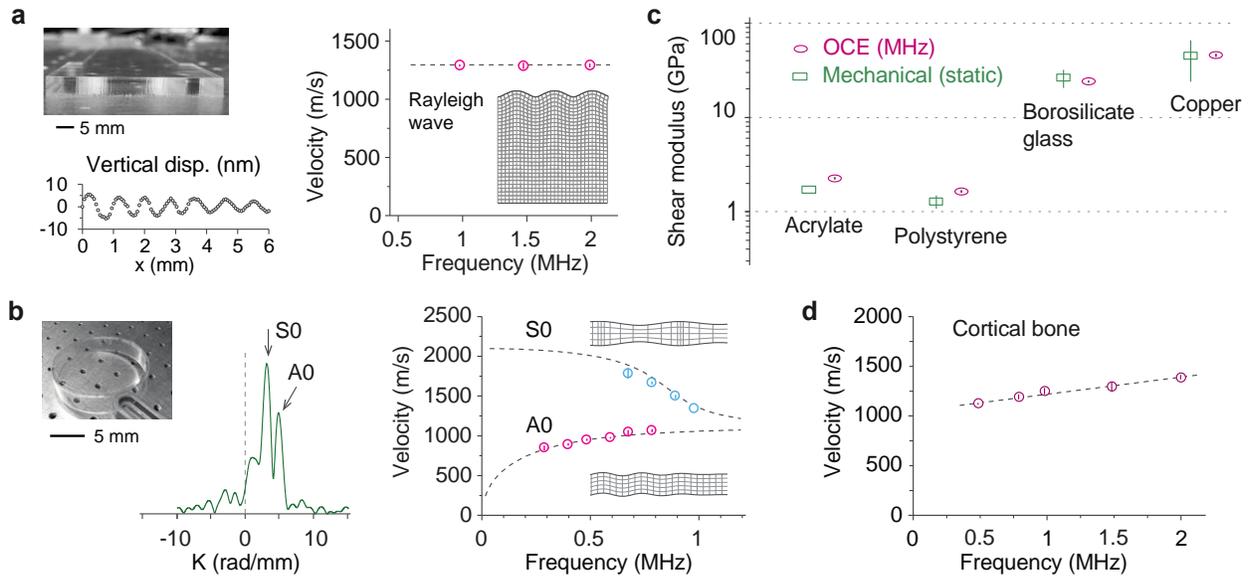

**Fig. 3 Ultrasonic OCE of hard materials. a**, Measured Rayleigh wave speeds (circles) at ultrasonic frequencies on an acrylate plastic block. Inset, a theoretical Rayleigh surface wave pattern. The measured surface displacement at 1 MHz is displayed. **b**, Measured asymmetric (A0) and symmetric (S0) Lamb wave modes on a 1 mm thick polystyrene bottom plate (circles) and the fitting result (dashed lines). Wavenumber domain plot displayed at 787.6 kHz resolves the two wave modes. **c**, Comparison of shear moduli measured by MHz OCE and literature values obtained from static mechanical testing. **d**, Measured wave speed on the cortical surface of a bovine tibia bone (circles) and the fitting result (dashed line). Vertical error bars represent the standard deviation of three measurements at different locations.

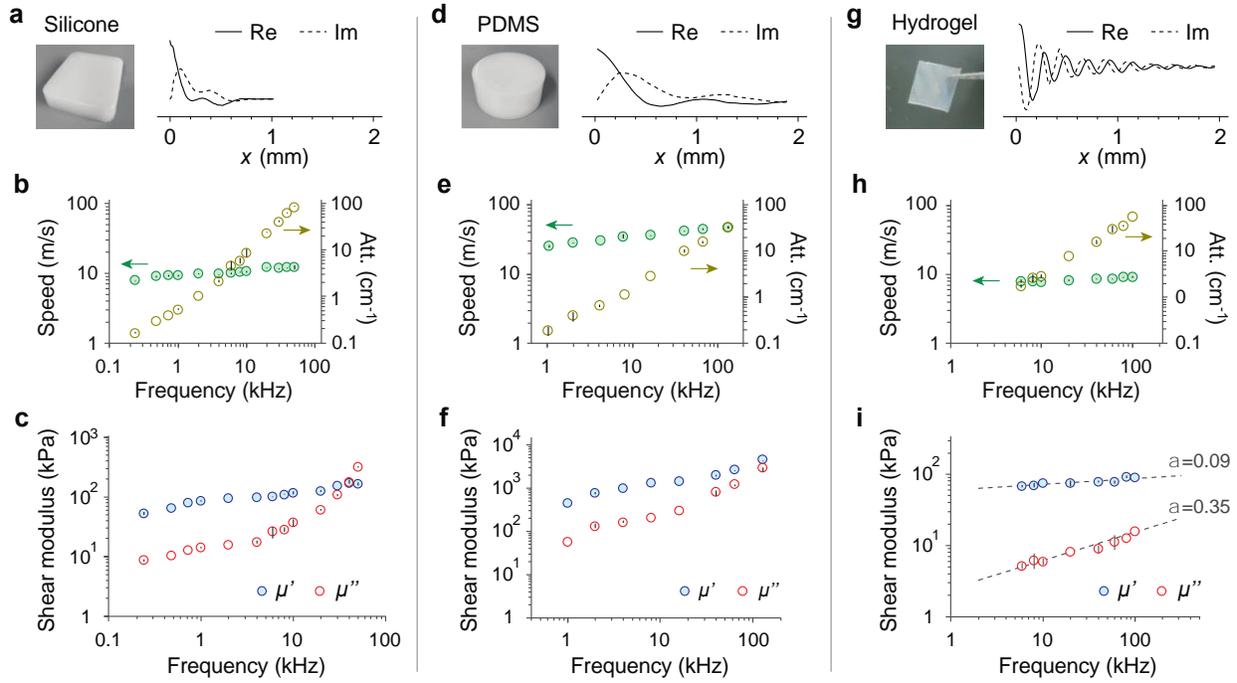

**Fig. 4: Ultra-wideband shear rheological analysis of uniform soft materials**: **a, b, c:** Rubber, **d, e, f**: PDMS, **g, h, i**: Hydrogel. **a**, Picture of the rubber and its surface wave displacement profile at 40 kHz (Re: real part, Im: imaginary part). **b**, Rubber's wave dispersion relation and attenuation. **c**, Rubber's storage modulus and loss modulus. **d**, Picture of the PDMS and its Rayleigh surface wave displacement profile at 40 kHz. **e**, PDMS's wave dispersion relation and attenuation. **f**, PDMS's storage modulus and loss modulus. **g**, Picture of the hydrogel and its Rayleigh surface wave displacement profile at 40 kHz. **h**, Hydrogel's wave dispersion relation and attenuation. **i**, Hydrogel's storage modulus and loss modulus. Data are represented as mean values ± std for measurements at three different locations. Dashed lines in (i) are curve fits using a power law $\mu \propto f^\alpha$.

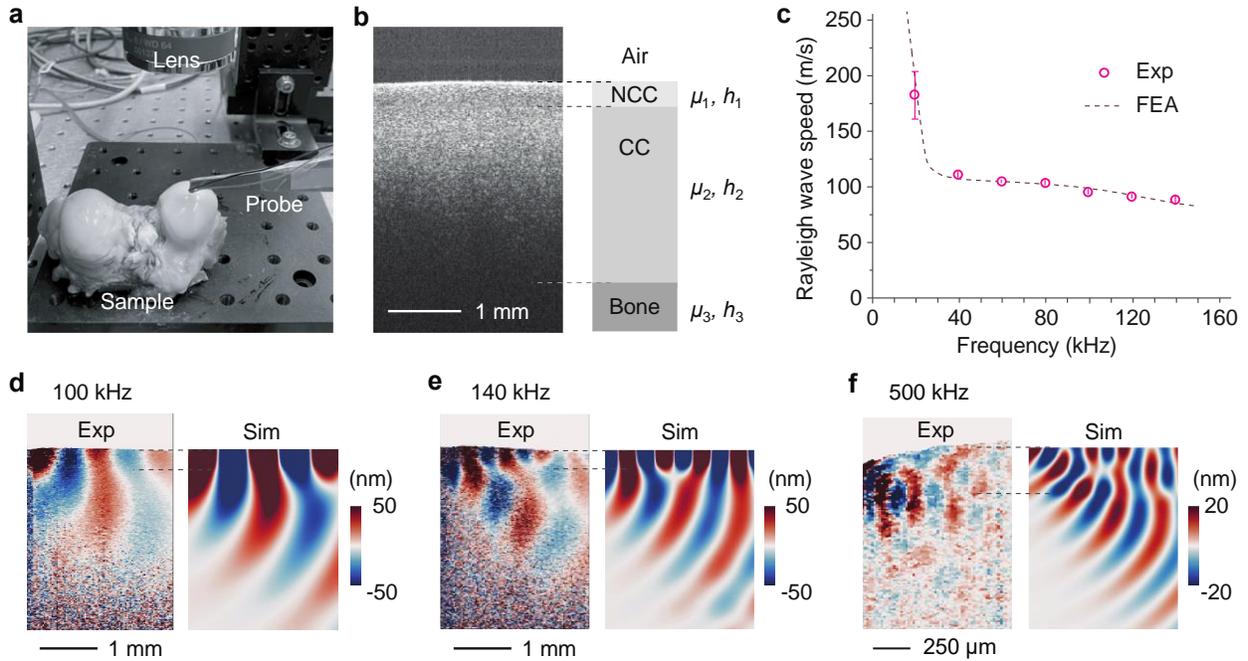

**Fig. 5 Depth-resolved stiffness mapping of joint tissue *ex vivo*. a**, Picture of a bovine joint sample and the PZT contact probe. **b**, Cross-sectional OCT image with three layers labeled: noncalcified cartilage (NCC), calcified cartilage (CC), and the bone. The shear moduli and thickness of the NCC, CC and bone layers are denoted by $\mu_i$ and $h_i$ ($i$ = 1, 2, 3), respectively. **c**, Measured Rayleigh surface wave speeds (circles) and the best-fit FEA result using the three-layer model (dashed curve). Error bars represent the standard deviation of four measurements from two samples at two locations. **d**-**f,** Side by side comparison of the measured cross-sectional displacement image and the FEA simulation at 100 kHz, 140 kHz, and 500 kHz, respectively. Dashed lines indicate the approximate boundaries of the NCC layer.

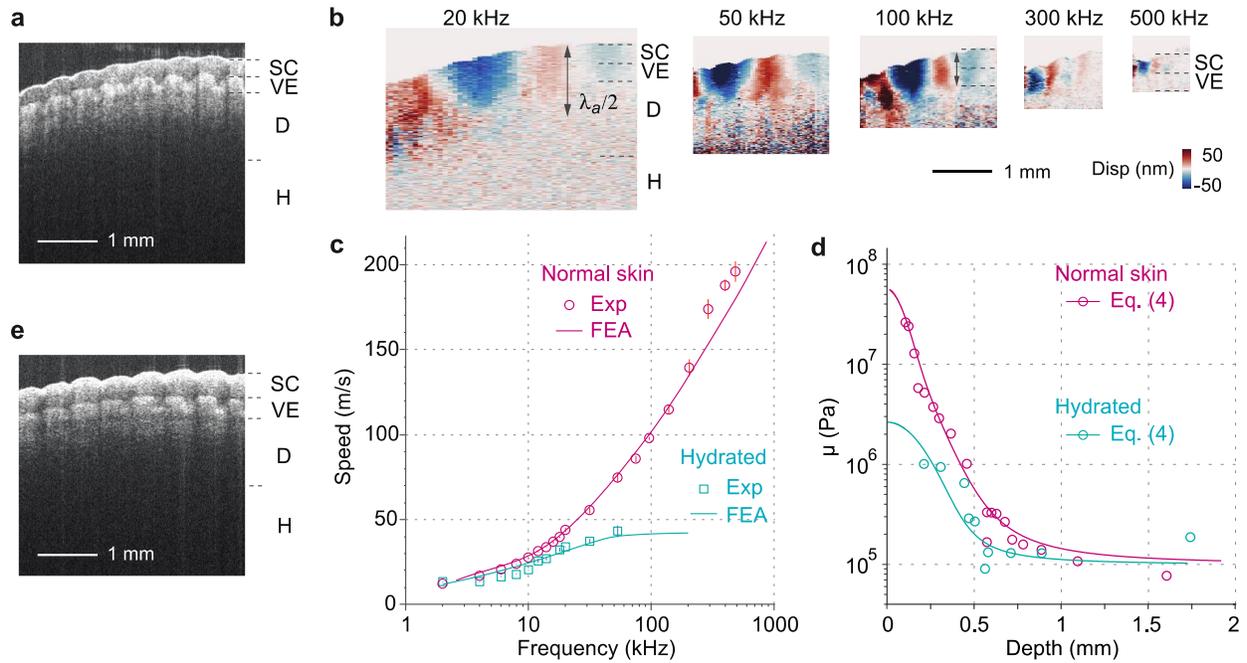

**Fig. 6 Depth-resolved stiffness mapping of human skin *in vivo*. a**, Cross-sectional OCT image of the fingertip skin in the normal condition. The stratum corneum (SC), viable epidermis (VE), dermis (D), and hypodermis (H) layers are labeled. **b**, Cross-sectional displacement images (only real part is shown) of normal fingertip at different frequencies. The layer interfaces and the half wavelength of the Rayleigh surface wave ($\lambda_a/2$) are marked. **c**, Measured Rayleigh surface wave speed over 2 – 500 kHz and the theoretical fitting curves for normal and hydration conditions. Error bars represent the standard deviation of three measurements from approximately same tissue regions. **d**, Calculated depth profiles of shear modulus for the normal and hydrated tissues. **e**, Cross-sectional OCT image of the same region after immersing in warm water for 20 min

# Supplementary Materials for

# Ultra-wideband optical coherence elastography from acoustic to ultrasonic frequencies

Xu Feng,[a,b,1] Guo-Yang Li,[a,b,1] and Seok-Hyun Yun[a,b,c],*

[a] Harvard Medical School and Wellman Center for Photomedicine, Massachusetts General Hospital, Boston, Massachusetts 02114, USA

[b] Department of Dermatology, Massachusetts General Hospital, Boston, Massachusetts 02114, USA

[c] Harvard-MIT Health Sciences and Technology, Cambridge, Massachusetts 02139, USA

[1] Co-first authors with equal contribution

* Correspondence to syun@hms.harvard.edu

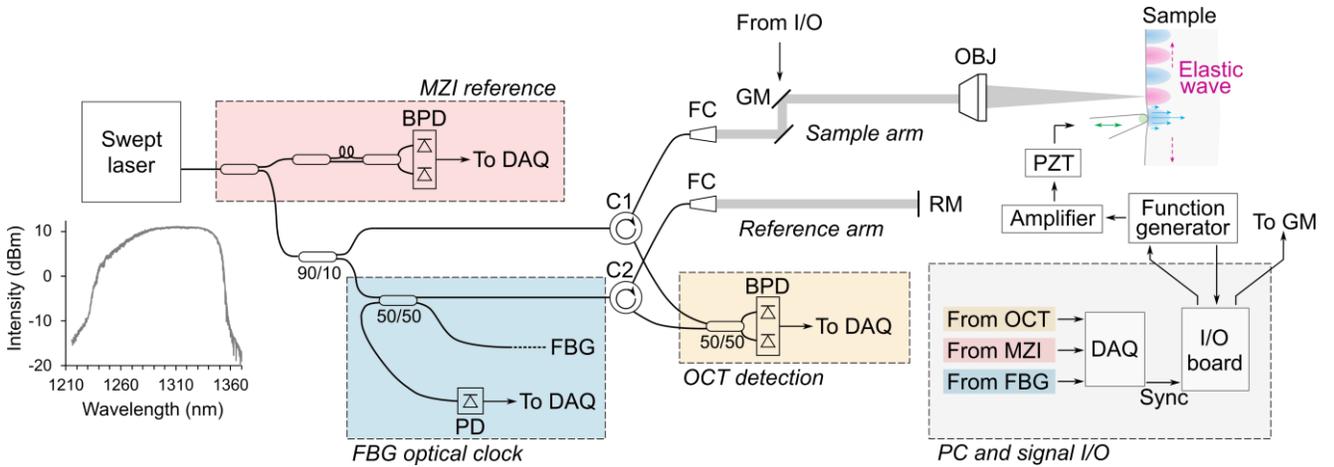

**Fig. S1. Schematic of the optical imaging system and signal generation and acquisition.** The output spectrum of the laser is measured by an optical spectral analyzer (AQ6370C, Yokogawa Electric). Abbreviations used: Mach-Zehnder interferometer (MZI), photodiode (PD), balanced photodiode (BPD), data acquisition (DAQ) system, Input/output (I/O) board, fiber Bragg grating (FBG),circulators (C1, C2), galvanometer mirrors (GM), fiber collimator (FC), objective lens (OBJ), reference mirror (RM), dichroic mirror (DM), piezoelectric transducer (PZT).

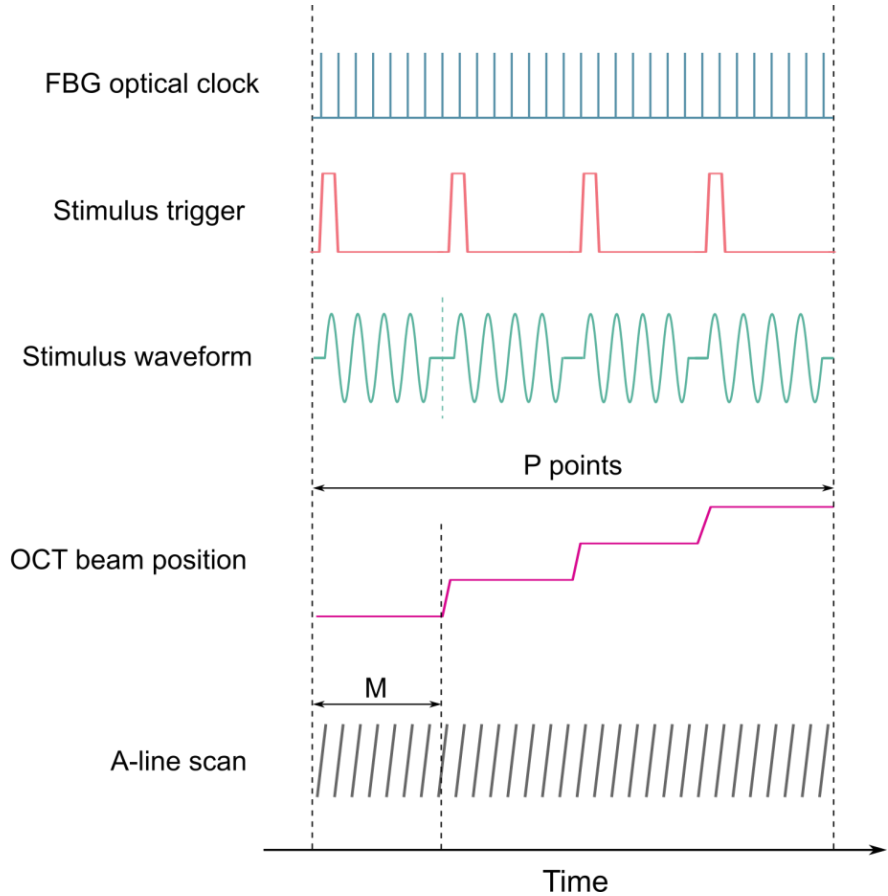

**Fig. S2. Timing diagrams of the OCE system.** The FBG optical clock provides the sync signal for the I/O board that controls the stimulus signal, the OCT beam scan, and data acquisition. For stimulus frequencies below 21.6 kHz, the I/O board generates the harmonic stimulus directly. For stimulus frequencies above 21.6 kHz, the I/O board generates a stimulus trigger to externally trigger a function generator to send the harmonic stimulus.

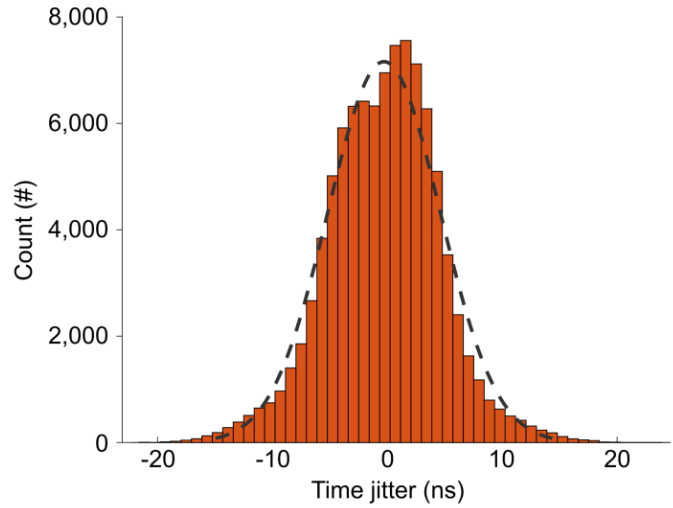

**Fig. S3. Gaussian distribution of the time jitter in sweep period** *T*. Data represents 900 M-scans, with each M-scan consisting of 108 A-lines. By evaluating the phase difference between two consecutive A-lines, we obtained the time jitter over 900*107 = 96,300 data points. The distribution is fitted with a Gaussian curve (dashed line). The standard deviation of the time jitter is 4.9 ns.

**Supplementary Note 1. Vibration noise introduced by laser time jitter**

According to Eq. (17) in the main text, the noise for the vibration phase primarily depends on the optical phase noise ($e^{-i2k_0z_0}$) and the laser timing jitter ($e^{-im\omega_m T}$). The latter one comes from the mechanical instability the sweep-source laser, which results in a timing jitter in the FBG clock, $T$. For our system, we get the standard deviation of $T$, denoted by $\sigma_T \approx 4.9$ ns.

Let's consider a measurement to a vibrating scatter using $N$ continuous A scan (M scan). The standard deviation of the vibration phase noise, denoted by $\sigma_\varphi$, is

$$\sigma_\varphi = \sqrt{(\sigma_{X'})^2 + (\sigma_L)^2}, \tag{S-24}$$

where $\sigma_L$ denotes the phase noise introduced by laser timing jitter and $\sigma_{X'}$ is the optical phase noise. $\sigma_{X'}$ is determined by the SNR of the sidelobe (denoted by $X'$),

$$\sigma_{X'} = \frac{1}{\sqrt{N}} \frac{1}{\sqrt{X'}}. \tag{S-25}$$

According to Eq. (16) in the main text,

$$10\log_{10}(X') = 10\log_{10}(X) + 20\log_{10}(k_0\delta), \tag{S-26}$$

where $X$ is the SNR of the static OCT signal (main lobe).

To derive the expression for $\sigma_L$, we denote the time interval of $m$-th A scan with $(T + t_m)$ ($m = 1,2,..,N$). Therefore, the mean and standard deviation of $t_m$ are 0 and $\sigma_T$, respectively. Then the time of the $m$-th A scan, denoted by $T_m$, is $T_m = \sum_{i=1}^{m}(T + t_i) = mT + \sum_{i=1}^{m} t_i$. The vibration signal, according to Eq. (17) in the main text, is $F_m \sim e^{i\omega_m T_m}$. We perform Fourier transform to this signal to get

$$\mathcal{F}(\omega) \sim \sum_{m=1}^{N} e^{i\omega_m T_m} e^{-\omega m T} = \sum_{m=1}^{N} e^{im(\omega_m - \omega)T} e^{i\omega_m (\sum_{i=1}^{m} t_i)}. \tag{S-27}$$

At $\omega = \omega_m$, we have

$$\mathcal{F}(\omega_m) \sim \sum_{m=1}^{N} e^{i\omega_m (\sum_{i=1}^{m} t_i)} \approx \sum_{m=1}^{N} [1 + i\omega_m (\sum_{i=1}^{m} t_i)]. \tag{S-28}$$

The phase of $\mathcal{F}(\omega_m)$, denoted by $\arg\{\mathcal{F}(\omega_m)\}$, is

$$\arg\{\mathcal{F}(\omega_m)\} \approx \frac{\omega_m \sum_{m=1}^{N}(\sum_{i=1}^{m} t_i)}{N} = \frac{\omega_m}{N}[Nt_1 + (N-1)t_2 + \cdots + t_N]. \tag{S-29}$$

Then $\sigma_L$ is the standard deviation of $\arg\{\mathcal{F}(\omega_m)\}$,

$$\sigma_L \approx \omega_m \sqrt{\left(\frac{N}{N}\right)^2 (\sigma_{t_1})^2 + \left(\frac{N-1}{N}\right)^2 (\sigma_{t_2})^2 + \cdots + \left(\frac{1}{N}\right)^2 (\sigma_{t_N})^2}. \tag{S-30}$$

where $\sigma_{t_m} = \sigma_T$ is the standard deviation of $t_m$ ($m = 1,2,..,N$). Therefore, we get

$$\sigma_L \approx \frac{\omega_m \sigma_T}{N} \sqrt{\frac{N(N+1)(2N+1)}{6}} \approx \sqrt{\frac{N}{3}}(\omega_m \sigma_T). \tag{S-31}$$

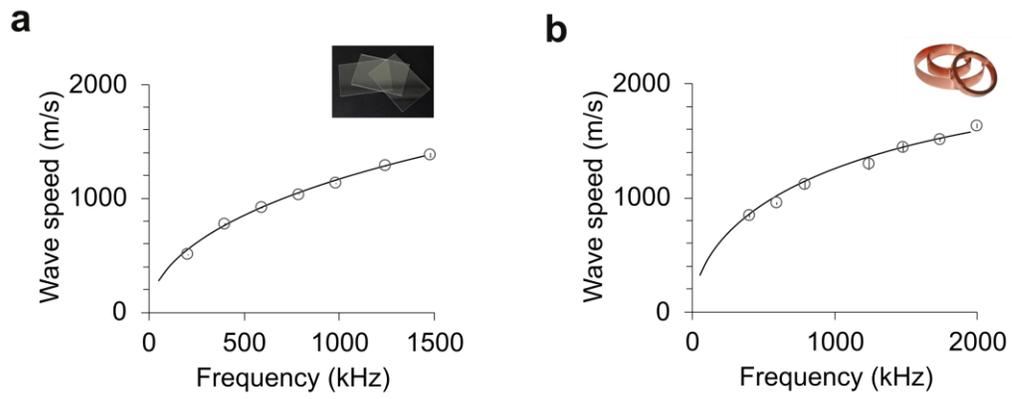

**Fig. S4. Ultrasonic OCE of the glass and metal. a**, Dispersion relation of the coverslip glass (dot) and theoretical fitting with a Lamb wave model (line). **b**, Dispersion relation of the copper sheet (dot) and the theoretical fitting with a Lamb wave model (line).

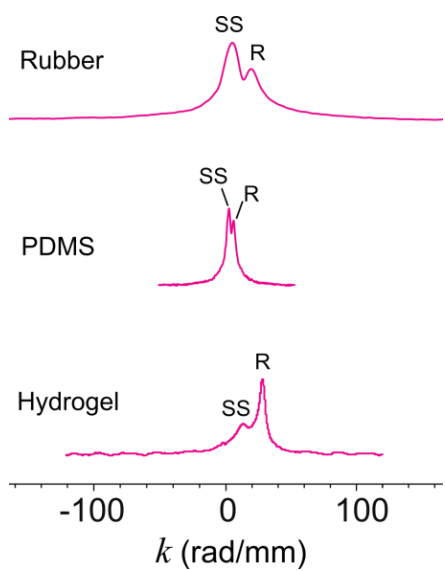

**Fig. S5.** Wavenumber k domain plots obtained by the Fourier Transform. Comparison is made at 40 kHz. The Rayleigh surface wave mode (R) and the supershear surface wave (SS) mode are labeled on the plots.

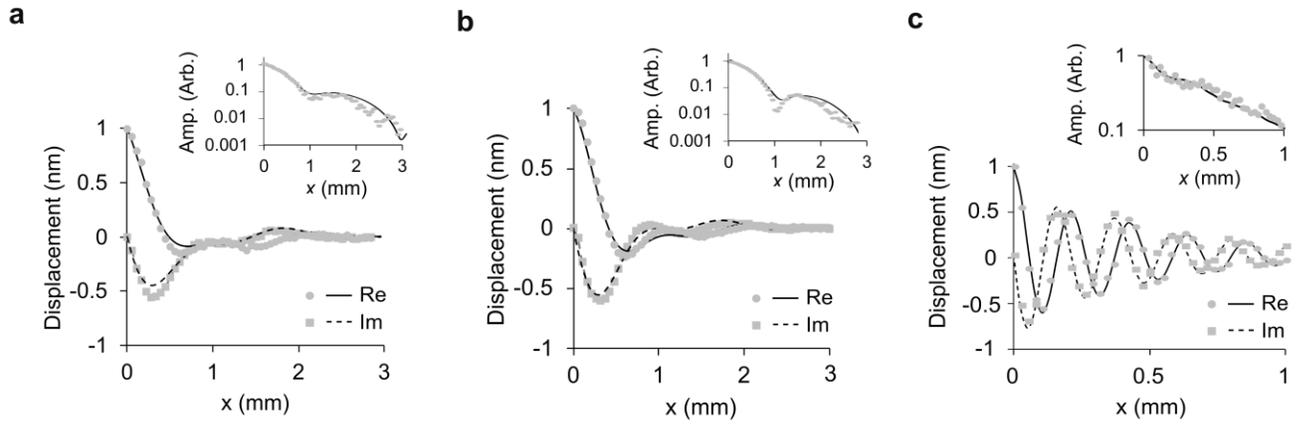

**Fig. S6. Rheological fitting of the soft materials. a**, Rubber (20 kHz). **b**, PDMS (40 kHz). **c**, Hydrogel (40 kHz). A comparison is maded between the experimental data and the theoretical model for the surface displacement. Markers, experiment; lines, theory. Inset: the total wave amplitude.

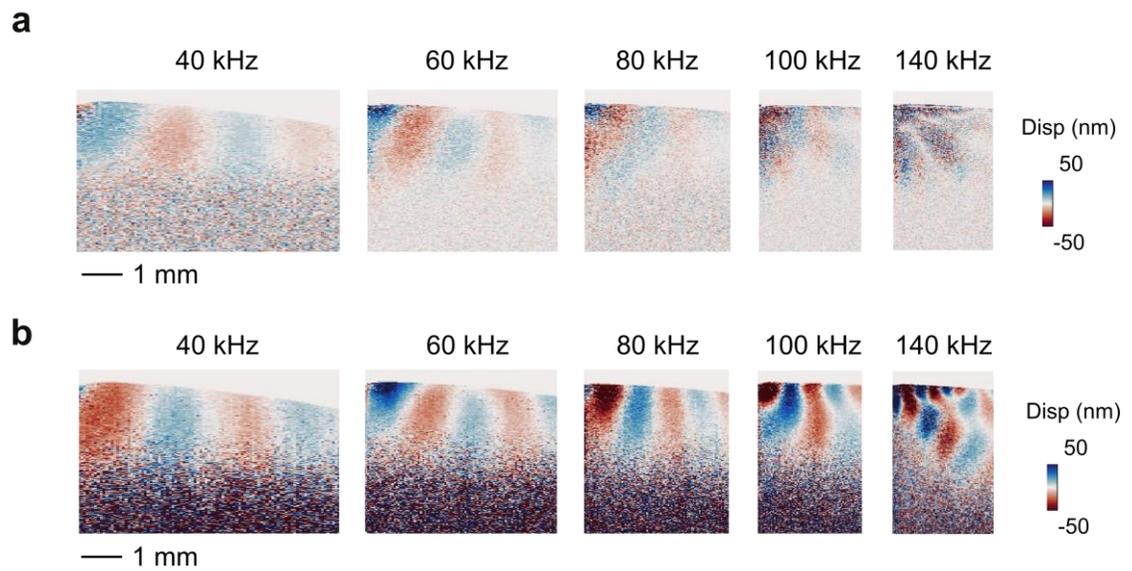

**Fig. S7. Cartilage OCE.** Cross-sectional displacement images at 40 – 140 kHz for **a**, Without applying the demodulation algorithm for inner scatters, and **b**, After applying the demodulation algorithm for inner scatterers.

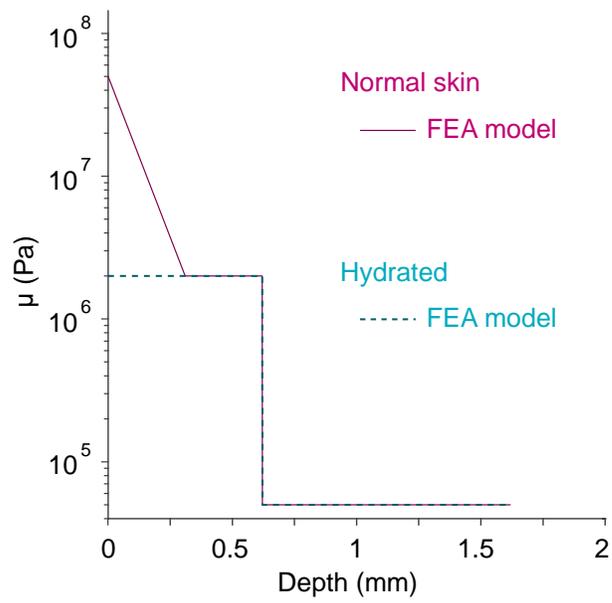

**Fig. S8. FEA model for the skin.** The depth dependent profile of shear modulus in the human skin for the normal, pre-hydrated state (margenta) and hydrated state (turquoise).

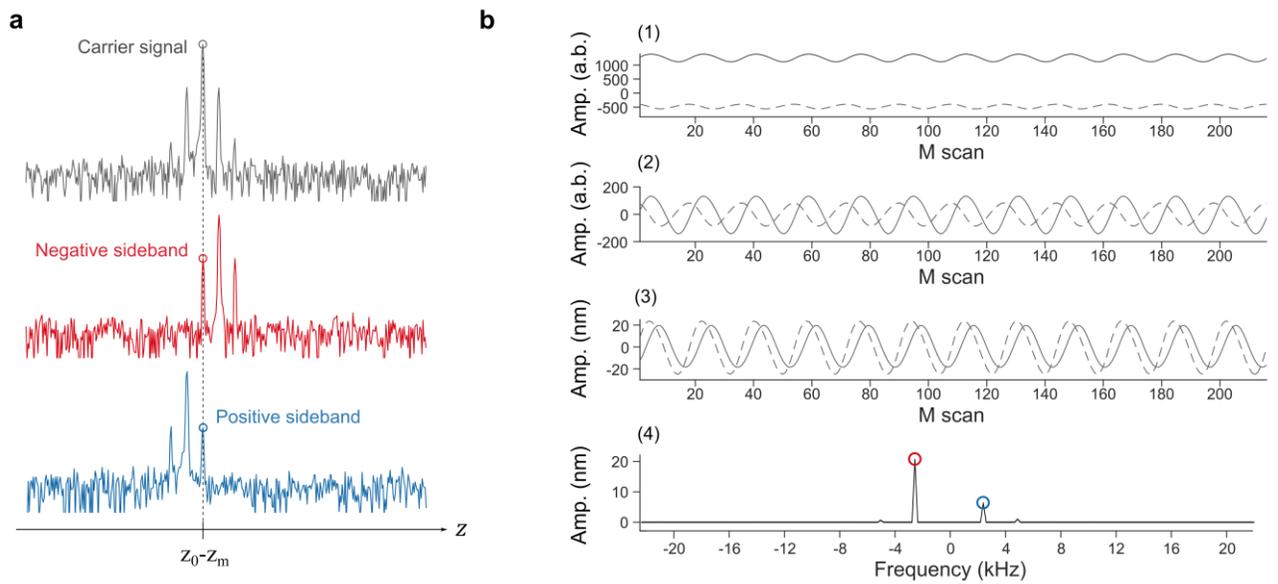

**Fig. S9. An example of signal demodulation. a,** The Fourier components at $F(\omega_0 - \omega_m)$ contain three signals: (1) The time-independent reflection signal (carrier signal) from a scatterer at $z_0 - z_m$ (top); (2) The vibration signal (negative sideband) originated from the scatterer at $\omega_0$ (middle); and (3) The vibration signal (positive sideband) originated from the scatterer located at $z_0 - 2z_m$ (bottom). **b,** Steps of signal demodulation. (1) Original signal $F(\omega_0 - \omega_m)$. (2) Subtracting the original signal from the background $\langle F(\omega_0 - \omega_m) \rangle$. (3) Normalizing the signal by $\langle F(\omega_0 - \omega_m) \rangle$. (4) Fourier transform to identify $\delta(z_0)$ and $\delta(z_0 - 2\omega_m)$.